# Hidden Thermodynamic Information in Protein Amino Acid Mutation Tables

V. Sachdeva and J. C. Phillips

Dept. of Physics and Astronomy, Rutgers University, Piscataway, N. J., 08854

**Abstract**

We combine the standard 1992 20x20 substitution matrix based on block alignment, BLOSUM62, with the standard 1982 amino acid hydropathicity scale KD as well as the modern 2007 hydropathicity scale MZ, and compare the results. The 20-parameter KD and MZ hydropathicity scales have different thermodynamic character, corresponding to first- and second-order transitions. The KD and MZ comparisons show that the mutation rates reflect quantitative iteration of qualitative amino acid –phobic and -philic binary 2x10 properties that define quaternary 4x5 subgroups (but not quinary 5x4 subgroups), with the modern MZ bioinformatic scale giving much better results. The quaternary 5-mer MZ 4x5 subgroups are called mutons (Mu5).

Protein amino acid sequences (aas) are rich in information, especially when combined with structural data. There are many Web-based tools for analyzing aas, but by far the most utilized is BLAST (**B**asic **L**ocal **A**lignment **S**earch **T**ool), which compares two given sequences, or searches for sequences similar to a given sequence. The original BLAST paper [1] was the most highly cited paper published in the 1990s. A key BLAST element is the "substitution matrix", which assigns a score for aligning any possible pair of residues, and identifies "positive" mutations between similar aas. The BLOSUM62 matrix (available online) is the default for most BLAST programs [2]. It obtains mutation rates $\Gamma$ of aa pairs from protein blocks (distantly related but conserved regions), which leads to accurate homological lists of functionally similar protein blocks.

Competing effects of hydrophobic and hydrophilic segments of a given protein have long been known to be the primary driving force behind the folding of protein chains into protein globules. There are secondary effects associated with longitudinal hydrogen bonding (α helices) and transverse hydrogen bonding (β strands), and even weaker charge effects, but in most proteins the dominant physico-chemical factor in a kinetic property such as aggregation [3] is hydropathic interactions.

Hydropathic interactions determine globular shapes and are manifested biochemically in many ways, which has led to many tables of aa hydropathicity Ψ. Here we will compare results obtained with the standard 1982 table, Ψ(KD) (17K citations)[4], and the modern 2007 Ψ(MZ) table, based on fractals and self-organized criticality [5]. The KD table is related to first-order effects (unfolding of globular proteins from water to air), while the MZ table describes second-order conformational changes in globular surface differential geometry [3]. Our analysis of the BLOSUM62 matrix will enable us to decide whether block homologies are primarily first- or second-order thermodynamically.

There is already a large literature on general aspects of biological evolution and statistical physics [6], which aim to go beyond phylogenetic trees based on point mutations (much less effective than blocks [2]). Explicit applications help to bring these general considerations into sharper perspective. The biomedically important area of viruses and vaccines is best quantified using epitopes [7,8], which are similar to but still different from blocks, which are best suited to describing self-sustaining proteins [9].

Mutational rates can be used to compare two aspects of different hydropathicity scales Ψ(aa), their discrete hierarchical ordering of 20 amino acids, with 20! possible orderings, or the continuum spacings between ordered amino acids. The MZ and KD orderings [5] can be divided into groups (hydropathic scale blocks), with the two obvious choices for replacing binary 2x10 =2x(2x5) by 4x5 = (2x2)x5, or quaternary 4 x 5aa blocks (denoted by Mut5), or quinary 5 x 4aa blocks (Mut4). The most hydrophobic Mut5α blocks, extracted from the BLOSUM62 matrix, are shown for the MZ and KD scales in Fig. 1. These groups could display the tendency of amino acids to mutate into other amino acids within their subgroup with similar hydropathicity.

One often sees qualitative comparisons of mutation rates of hydro (phobic,philic) aa, but with Mut groups one can make quantitative comparisons. One averages the mutational off-diagonal group matrix elements, and compares those averages with the hydropathic width of each group (defined as Ψ(first aa) - Ψ(last aa)). The wider the subgroup, the more Ψ phase space is available for internal mutations. This idea can be tested at the simple 2x10 hydrophobic/hydrophilic level, for Ω = (Mut4α + Mut4β) - (Mut4γ + Mut4δ). With the MZ scale Ω = -1.2 (as one might have expected, exposed hydrophilic aa mutate much more often than buried hydrophobic aa), but with the KD scale Ω = 0.2 (unsatisfactory).

This binary phobic/philic test is coarse: what happens when we calculate the Mut5 correlations of average mutation rates with average hydropathic widths? The results are R = 0.93 (MZ) and 0.86 (KD), both very successful, but MZ is even more successful. When we repeat these steps with the Mut4 groups, we obtain weak and inconsistent results: R = 0.3 (MZ) and -0.3 (KD). The 4x5 Mut5 partition is an iteration of the 2x10 phobic/philic partition, which explains its block success, as well as the failure of the non-iterated Mut4 blocks.

The symmetry of BLOSUM mutation rates suggests that we examine successive waves (generations) of mutation rates, corresponding to diagonal strings parallel to the principal (unmutated) diagonal. Again we reduce noise by looking at average rates Φ(N) of groups, but now the groupings Λ(N, N + κ) are averaged over waves based on the hydropathically ordered aa sequences (in matrix terms, Γ elements (N, N + κ) are averaged over κ, from κ = 1 to $\kappa_{max}$). The results using the MD ordering are shown in Fig. 2, and the KD ordering in Fig. 3, and discussed in those captions.

A striking feature of the mutational waves is the smoothness of the MZ Φ(N) groups (Fig. 2) compared to the KD Φ(N) groups (Fig.3). Here we define roughness as the 20-aa average over N of $(\Phi(N) - \Phi(N+1))^2$. This roughness is presumably a measure of the thermodynamic noise of block mutation rates, averaged over thousands of proteins. As shown in Fig. 4, this noise is almost the same for the MZ and KD scales for $\kappa_{max} \geq 10$, but for $\kappa_{max} = 5$, MZ is 35% smoother, presumably reflecting a greater information content using the fractal MZ conformational scale, compared to the unfolding KD scale. Note that it has been found that ~ 4 mutations within epitope A or B, between the old vaccine target strain and the currently dominating circulating strain, are enough to render the H3N2 vaccine ineffective [7-9]. This is a high level of internal consistency between two widely separated methods.

Broad averages minimize protein differences. GenomeNet, a Web-based index of protein scales, lists 34 hydropathicity scales from the classic period 1968-1995 [10]. The present results show that all hydropathicity scales are not equal. The modern 2007 MZ fractal scale is much more accurate than the standard classic KD scale in describing the evolution of lysozyme *c* [3]. Similarly epitope analysis and dimensional compression are more accurate than phylogenetic trees in identifying clusters of rapidly drifting viral strains, the critical factor in engineering effective H3N2 vaccines [7-9] and arXiv 1510.00488.

Spatial modeling suggests a simple picture for the universal fractal properties discovered by MZ [5]. The first popular application of hydropathicity scales was to the study of all-α heptad transmembrane opsins, whose seven internal transmembrane segments are predominantly hydrophobic, with a typical length around 20 amino acids. One can carry this description to the similarly 20 amino acid thick layers adjacent to cell membranes where proteins can interact most effectively, as they are temporarily confined to a narrower space, either with transmembrane loops, or with other proteins. In this frontier surface layer space proteins can both interact on a 20 amino acid length scale, or evolve via exchange of modular elements of ~20 amino acids. This picture is analogous to heterogeneous catalysis, with the membrane playing the part of catalytic substrate.

Our present analysis has shown a deeper and more specific function of protein amino acid chemistry, connected with iteration of the phobic/philic partition, which may catalytically accelerate adaptive evolution. This underlying sub-membrane motif, the Mut5 partitioning of the fractal MZ scale, is nearly universal, according to the success of BLOSUM62. Thus the MZ Mut5 5-mers could be called mutons. Starting from the most hydrophobic, these can be labelled α, β, γ and δ. The best-known group here is the chargeable group δ, which includes Lys,Arg, Asp and Glu (K,R, D and E). An interesting question is what is the fifth amino acid in this group: on the KD scale it is Asn (N), while on the MZ scale it is Ser (S). Elsewhere this question is discussed in detail, for analyzing the evolution of vaccine effectiveness for flu H3N2 [11,12]

The matrix manipulations described here may seem puzzling to biologists unfamiliar with matrix algebra, but they should present few challenges to physicists familiar with quantum matrix algebra. They enable the compression of very large quantities of biological data, simply by looking for features that are described more accurately with the MZ scale and mutons. Many classic concepts could be refined in this way. For example, one could study some of the eight examples of protein antigenic determinants discussed long ago by Hopp and Woods (3400 citations), using a crude hydropathicity scale [13]. In the intervening decades, there has been substantial progress in each case, especially hemagglutinin [7-9], but by tracing the histories using the comparative advantages of the modern fractal MZ scale, new insights are easily obtained. This procedure is greatly facilitated by using the Web of Science, a widely underutilized tool [14].

The advantages of using a more accurate hydropathicity scale have not been recognized in the biological literature. Since 2007, the classical 1982 KD Ψ scale has been cited > 3300 times, while the modern MZ Ψ scale has been cited only 1% as often. The differences between the two Ψ scales, although small, lead not only to the large universal muton effects discussed here, but even larger differences for specific proteins [3]. In specific applications, modern bioinformatic scaling methods can advance the development of biomedically important subjects, such as a simple blood test for early cancer detection (arXiv 1509.01577). The results presented here show that most mutations are more nearly small second-order effects, than large first-order effects. For the reader's convenience, the MZ and KD reordered BLOSUM62 matrices, as well as the MZ and KD diagonal mutation wave tables, are attached to this paper at arXiv xxx. The MZ and KD Ψ tables are in [15].

| | | | | |
|---|---|---|---|---|
| | -1 | -1 | -2 | -1 |
| | | 3 | -1 | 1 |
| MZ | | | -1 | 1 |
| α = CVI YM | | | | -1 |
| | 3 | 2 | 0 | -1 |
| | | 1 | -1 | -1 |
| KD | | | 0 | -1 |
| α = IVLFC | | | | -2 |

Fig. 1. Two sample Mut5 blocks, using either the MZ or KD hydropathicity scale. The integers represent logs of mutation rates, in (bit2)/2 units. Only the off-diagonal elements are shown, and the MZα rows (columns) are labelled by VIYM (CVIY). For the MZ group α, the hydropathic width is Ψ(C) - Ψ(M). The KD block is similar, but with α = IVLFC.

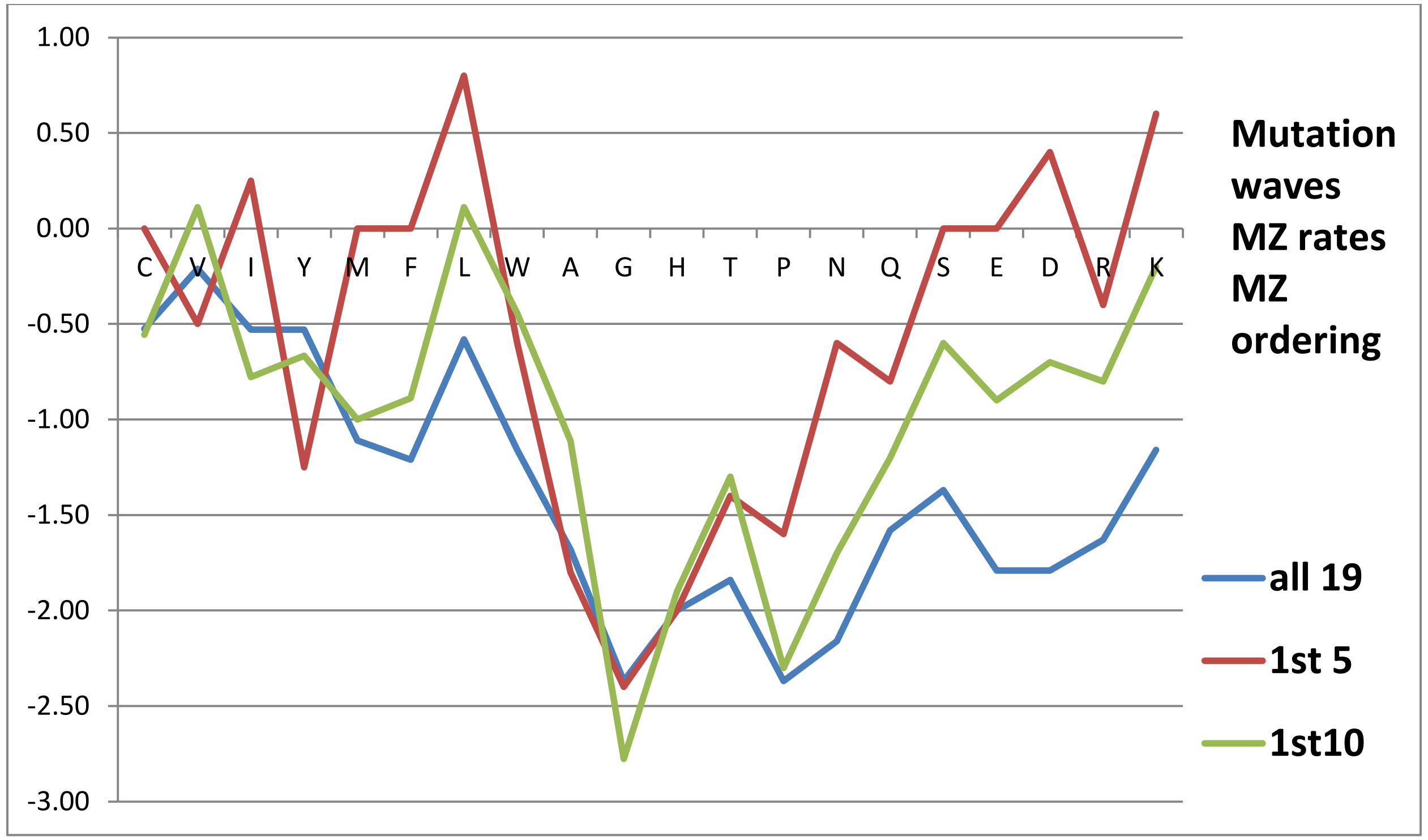

Fig. 2. When the MZ scale is used to construct mutation waves Λ(N, N + κ), a strong dip in averaged wave rates Φ(N) is seen near the hydroneutral center aa. The hydrophobic and hydrophilic wings have higher mutational rates. Here $1^{st}$ 5 means Φ(N) is obtained by averaging Λ(N, N + κ) over κ = 1 to 5,etc. The memory effects for larger κ are large for the hydroneutral dip, and small for the hydrophilic wing.

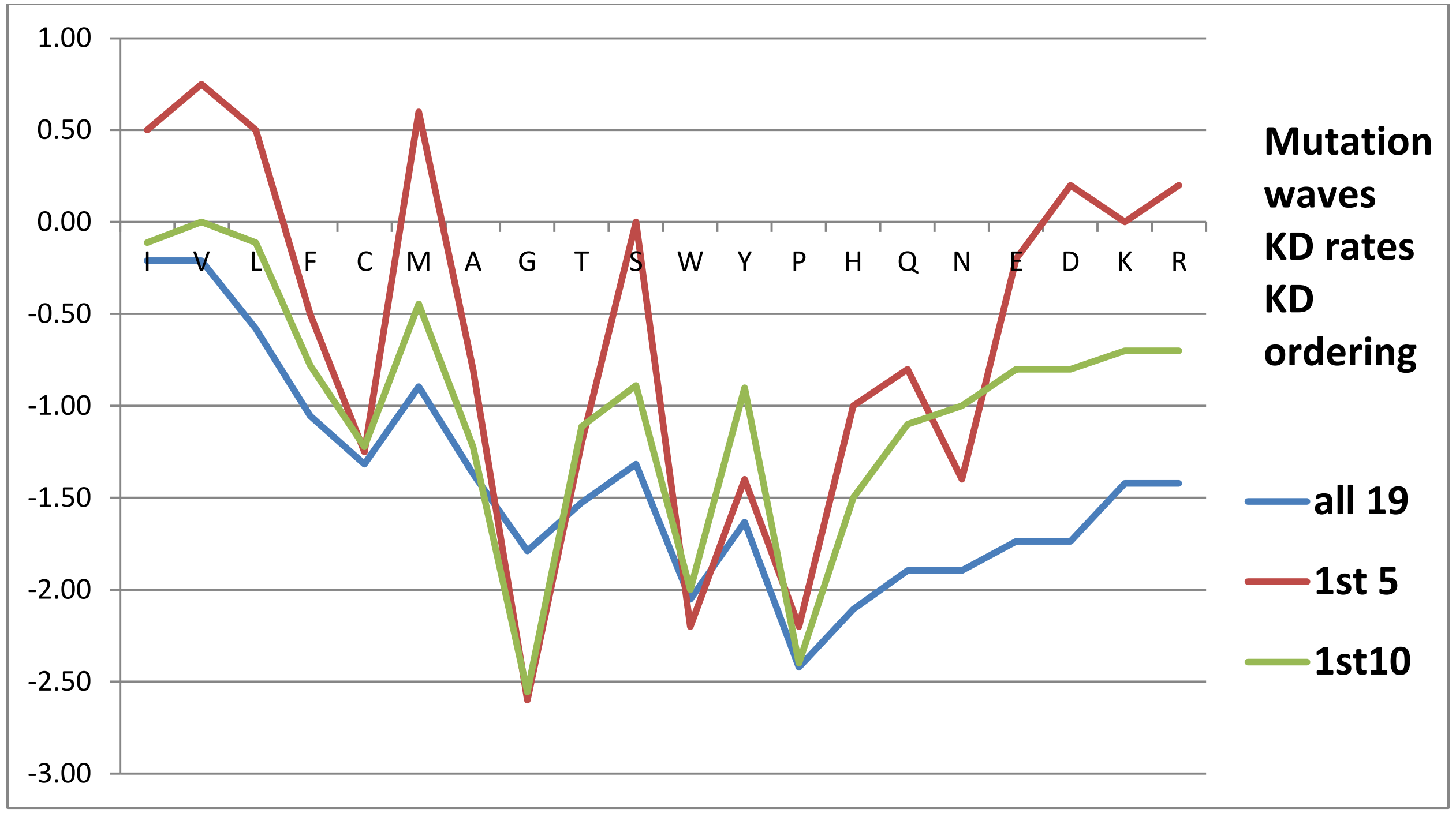

Fig. 3. This is similar to Fig. 2, except that here the KD hierarchy has been used to order the BLOSUM62 matrix [2]. Much of the simplicity of Fig. 2 has been lost, presumably because mutational substitutions tend to minimize MZ second-order conformational distortions, and do not involve the first-order unfolding effects described by the KD scale.

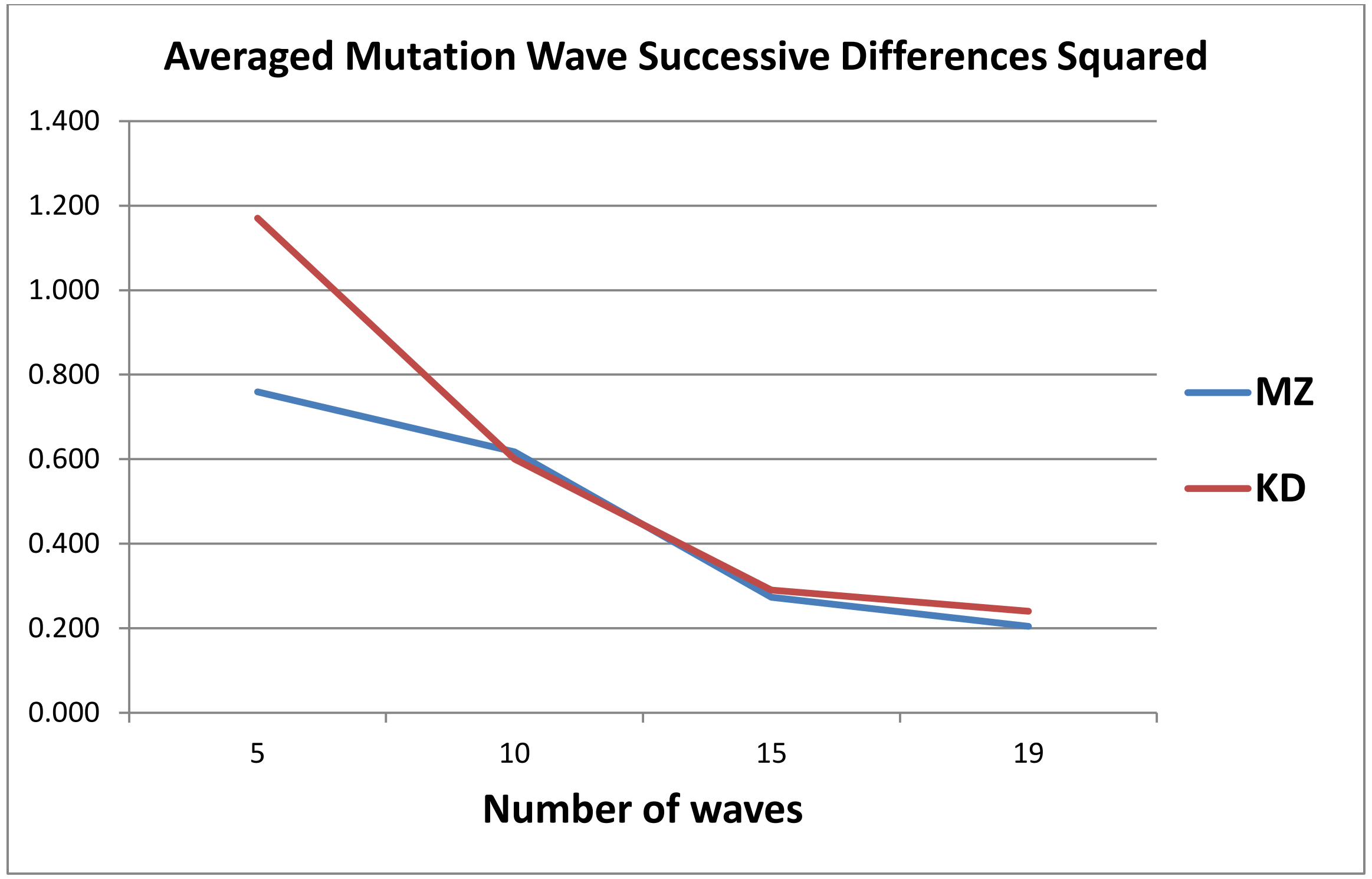

Fig. 4. Quantitative comparison of Φ(N) from Figs. 3 and 4 roughness Ω, defined as average of $(\Phi(N) - \Phi(N+1))^2$. The two orderings yield almost identical Ω for $\kappa_{max} \geq 10$, but for $\kappa_{max} = 5$, the MZ ordering yields 35% smoother successive mutation rate waves.

| | C | V | I | Y | M | F | L | W | A | G | H | T | P | N | Q | S | E | D | R | K |
|---|---|---|---|---|---|---|---|---|---|---|---|---|---|---|---|---|---|---|---|---|
| C | 9 | -1 | -1 | -2 | -1 | -2 | -1 | -2 | -3 | -3 | -3 | -1 | -3 | -3 | -3 | -1 | -4 | -3 | -3 | -3 |
| V | | 4 | 3 | -1 | 1 | -1 | 1 | -3 | 0 | -3 | -3 | 0 | -2 | -3 | -2 | -2 | -2 | -3 | -3 | -2 |
| I | | | 4 | -1 | 1 | 0 | 2 | -3 | -1 | -4 | -3 | -1 | -3 | -3 | -3 | -2 | -3 | -3 | -3 | -3 |
| Y | | | | 7 | -1 | 3 | -1 | 2 | -2 | -3 | 2 | -2 | -3 | -2 | -1 | -2 | -2 | -3 | -2 | -2 |
| M | | | | | 5 | 0 | 2 | -1 | -1 | -3 | -2 | -1 | -2 | -2 | 0 | -1 | -2 | -3 | -1 | -1 |
| F | | | | | | 6 | 0 | 1 | -2 | -3 | -1 | -2 | -4 | -3 | -3 | -2 | -3 | -3 | -3 | -3 |
| L | | | | | | | 4 | -2 | -1 | -4 | -3 | -1 | -3 | -3 | -2 | -2 | -3 | -4 | -2 | -2 |
| W | | | | | | | | 11 | -3 | -2 | -2 | -2 | -4 | -4 | -2 | -3 | -3 | -4 | -3 | -3 |
| A | | | | | | | | | 4 | 0 | -2 | 0 | -1 | -2 | -1 | 1 | -1 | -2 | -1 | -1 |
| G | | | | | | | | | | 6 | -2 | -2 | 0 | 0 | -2 | 0 | -2 | -1 | -2 | -2 |
| H | | | | | | | | | | | 8 | -2 | -2 | 1 | 0 | -1 | 0 | -1 | 0 | -1 |
| T | | | | | | | | | | | | 5 | -1 | 0 | -1 | 1 | -1 | -1 | -1 | -1 |
| P | | | | | | | | | | | | | 7 | -2 | -1 | -1 | -1 | -1 | -2 | -1 |
| N | | | | | | | | | | | | | | 6 | 0 | 1 | 0 | 1 | 0 | 0 |
| Q | | | | | | | | | | | | | | | 5 | 0 | 2 | 0 | 1 | 1 |
| S | | | | | | | | | | | | | | | | 4 | 0 | 0 | -1 | 0 |
| E | | | | | | | | | | | | | | | | | 5 | 2 | 0 | 1 |
| D | | | | | | | | | | | | | | | | | | 6 | -2 | -1 |

**MZ Table**

| | | | | | | | | | | | | | | | | | | | | |
|---|---|---|---|---|---|---|---|---|---|---|---|---|---|---|---|---|---|---|---|---|
| R | | | | | | | | | | | | | | | | | | | 5 | 2 |
| K | | | | | | | | | | | | | | | | | | | | 5 |

| **MZ waves** | 1 | 2 | 3 | 4 | 5 | 6 | 7 | 8 | 9 | 10 | 11 | 12 | 13 | 14 | 15 | 16 | 17 | 18 | 19 | 20 |
|---|---|---|---|---|---|---|---|---|---|---|---|---|---|---|---|---|---|---|---|---|
| | C | V | I | Y | M | F | L | W | A | G | H | T | P | N | Q | S | E | D | R | K |
| 1 | | -1 | 3 | -1 | -1 | 0 | 0 | -2 | -3 | 0 | -2 | -2 | -1 | -2 | 0 | 0 | 0 | 2 | -2 | 2 |
| 2 | -1 | | -1 | -1 | 1 | 3 | 2 | 1 | -1 | -2 | -2 | -2 | -2 | 0 | -1 | 1 | 2 | 0 | 0 | -1 |
| 3 | 3 | -1 | | -2 | 1 | 0 | -1 | -1 | -2 | -4 | -2 | 0 | 0 | 1 | -1 | -1 | 0 | 0 | -1 | 1 |
| 4 | -1 | -1 | -2 | | -1 | -1 | 2 | 2 | -1 | -3 | -3 | -2 | -1 | 0 | 0 | 1 | -1 | 1 | 1 | 0 |
| 5 | -1 | 1 | 1 | -1 | | -2 | 1 | -3 | -2 | -3 | -1 | -1 | -4 | -2 | -2 | -1 | -1 | -1 | 0 | 1 |
| 6 | 0 | 3 | 0 | -1 | -2 | | -1 | -3 | -1 | -3 | -2 | -2 | -3 | -4 | -1 | 0 | 0 | -1 | -2 | 0 |
| 7 | 0 | 2 | -1 | 2 | 1 | -1 | | 2 | 0 | -4 | 2 | -1 | -4 | -3 | -2 | 1 | -2 | -1 | -1 | -1 |
| 8 | -2 | 1 | -1 | 2 | -3 | -3 | 2 | | 3 | -3 | -3 | -2 | -2 | -3 | -2 | -3 | -1 | -1 | 0 | -1 |
| 9 | -3 | -1 | -2 | -1 | -2 | -1 | 0 | 3 | | -3 | -3 | -1 | -3 | -2 | -3 | -2 | -3 | -2 | -2 | -1 |
| 10 | 0 | -2 | -4 | -3 | -3 | -3 | -4 | -3 | -3 | | -3 | 0 | -3 | -2 | 0 | -2 | -3 | -4 | -1 | -2 |
| 11 | -2 | -2 | -2 | -3 | -1 | -2 | 2 | -3 | -3 | -3 | | -1 | -2 | -3 | -1 | -1 | -3 | -4 | -3 | -1 |
| 12 | -2 | -2 | 0 | -2 | -1 | -2 | -1 | -2 | -1 | 0 | -1 | | -3 | -3 | -3 | -2 | -2 | -3 | -2 | -3 |
| 13 | -1 | -2 | 0 | -1 | -4 | -3 | -4 | -2 | -3 | -3 | -2 | -3 | | -3 | -2 | -2 | -2 | -3 | -3 | -2 |
| 14 | -2 | 0 | 1 | 0 | -2 | -4 | -3 | -3 | -2 | -2 | -3 | -3 | -3 | | -3 | -2 | -3 | -3 | -1 | -3 |
| 15 | 0 | -1 | -1 | 0 | -2 | -1 | -2 | -2 | -3 | 0 | -1 | -3 | -2 | -3 | | -1 | -2 | -3 | -2 | -1 |
| 16 | 0 | 1 | -1 | 1 | -1 | 0 | 1 | -3 | -2 | -2 | -1 | -2 | -2 | -2 | -1 | | -4 | -3 | -3 | -2 |
| 17 | 0 | 2 | 0 | -1 | -1 | 0 | -2 | -1 | -3 | -3 | -3 | -2 | -2 | -3 | -2 | -4 | | -3 | -3 | -3 |
| 18 | 2 | 0 | 0 | 1 | -1 | -1 | -1 | -1 | -2 | -4 | -4 | -3 | -3 | -3 | -3 | -3 | -3 | | -3 | -2 |
| 19 | -2 | 0 | -1 | 1 | 0 | -2 | -1 | 0 | -2 | -1 | -3 | -2 | -3 | -1 | -2 | -3 | -3 | -3 | | -3 |
| 20 | 2 | -1 | 1 | 0 | 1 | 0 | -1 | -1 | -1 | -2 | -1 | -3 | -2 | -3 | -1 | -2 | -3 | -2 | -3 | |

| **KD Table** | I | V | L | F | C | M | A | G | T | S | W | Y | P | H | Q | N | E | D | K | R |
|---|---|---|---|---|---|---|---|---|---|---|---|---|---|---|---|---|---|---|---|---|
| I | 4 | 3 | 2 | 0 | -1 | 1 | -1 | -4 | -1 | -2 | -3 | -1 | -3 | -3 | -3 | -3 | -3 | -3 | -3 | -3 |
| V | 3 | 4 | 1 | -1 | -1 | 1 | 0 | -3 | 0 | -2 | -3 | -1 | -2 | -3 | -2 | -3 | -2 | -3 | -2 | -3 |
| L | 2 | 1 | 4 | 0 | -1 | 2 | -1 | -4 | -1 | -2 | -2 | -1 | -3 | -3 | -2 | -3 | -3 | -4 | -2 | -2 |
| F | 0 | -1 | 0 | 6 | -2 | 0 | -2 | -3 | -2 | -2 | 1 | 3 | -4 | -1 | -3 | -3 | -3 | -3 | -3 | -3 |
| C | -1 | -1 | -1 | -2 | 9 | -1 | 0 | -3 | -1 | -1 | -2 | -2 | -3 | -3 | -3 | -3 | -4 | -3 | -3 | -3 |
| M | 1 | 1 | 2 | 0 | -1 | 5 | -1 | -3 | -1 | -1 | -1 | -1 | -2 | -2 | 0 | -2 | -2 | -3 | -1 | -1 |
| A | -1 | 0 | -1 | -2 | 0 | -1 | 4 | 0 | 0 | 1 | -3 | -2 | -1 | -2 | -1 | -2 | -1 | -2 | -1 | -2 |
| G | -4 | -3 | -4 | -3 | -3 | -3 | 0 | 6 | -2 | 0 | -2 | -3 | -2 | -2 | -2 | 0 | -2 | -1 | -2 | -2 |
| T | -1 | 0 | -1 | -2 | -1 | -1 | 0 | -2 | 5 | 1 | -2 | -2 | -1 | -2 | -1 | 0 | -1 | -1 | -1 | -1 |
| S | -2 | -2 | -2 | -2 | -1 | -1 | 1 | 0 | 1 | 4 | -3 | -2 | -1 | -1 | 0 | 1 | 0 | 0 | 0 | -1 |
| W | -3 | -3 | -2 | 1 | -2 | -1 | -3 | -2 | -2 | -3 | 11 | 2 | -4 | -2 | -2 | -4 | -3 | -4 | -3 | -3 |
| Y | -1 | -1 | -1 | 3 | -2 | -1 | -2 | -3 | -2 | -2 | 2 | 7 | -3 | 2 | -1 | -2 | -2 | -3 | -2 | -2 |
| P | -3 | -2 | -3 | -4 | -3 | -2 | -1 | -2 | -1 | -1 | -4 | -4 | 7 | -2 | -1 | -2 | -1 | -1 | -1 | -2 |
| H | -3 | -3 | -3 | -1 | -3 | -2 | -2 | -2 | -2 | -1 | -2 | -2 | -2 | 8 | 0 | 1 | 0 | -1 | -1 | 0 |
| Q | -3 | -2 | -2 | -3 | -3 | 0 | -1 | -2 | -1 | 0 | -2 | -2 | -1 | 0 | 5 | 0 | 2 | 0 | 1 | 1 |
| N | -3 | -3 | -3 | -3 | -3 | -2 | -2 | 0 | 0 | 1 | -4 | -4 | -2 | 1 | 0 | 6 | 0 | 1 | 0 | 0 |
| E | -3 | -2 | -3 | -3 | -4 | -2 | -1 | -2 | -1 | 0 | -3 | -3 | -1 | 0 | 2 | 0 | 5 | 2 | 1 | 0 |
| D | -3 | -3 | -4 | -3 | -3 | -3 | -2 | -1 | -1 | 0 | -4 | -4 | -1 | -1 | 0 | 1 | 2 | 6 | -1 | -2 |
| K | -3 | -2 | -2 | -3 | -3 | -1 | -1 | -2 | -1 | 0 | -3 | -3 | -1 | -1 | 1 | 0 | 1 | -1 | 5 | 2 |
| R | -3 | -3 | -2 | -3 | -3 | -1 | -2 | -2 | -1 | -1 | -3 | -3 | -2 | 0 | 1 | 0 | 0 | -2 | 2 | 5 |

| **KD Waves** | 1 | 2 | 3 | 4 | 5 | 6 | 7 | 8 | 9 | 10 | 11 | 12 | 13 | 14 | 15 | 16 | 17 | 18 | 19 | 20 |
|---|---|---|---|---|---|---|---|---|---|---|---|---|---|---|---|---|---|---|---|---|
| | I | V | L | F | C | M | A | G | T | S | W | Y | P | H | Q | N | E | D | K | R |
| 1 | | 3 | 1 | 0 | -2 | -1 | -1 | 0 | -2 | 1 | -3 | 2 | -3 | -2 | 0 | 0 | 0 | 2 | -1 | 2 |
| 2 | 3 | | 2 | -1 | -1 | 0 | 0 | -3 | 0 | 0 | -2 | -2 | -4 | 2 | -1 | 1 | 2 | 1 | 1 | -2 |
| 3 | 1 | 2 | | 0 | -1 | 2 | -2 | -3 | -1 | 1 | -2 | -2 | -1 | -2 | -1 | -2 | 0 | 0 | 0 | 0 |
| 4 | 0 | -1 | 0 | | -1 | 1 | -1 | -3 | -1 | -1 | -3 | -3 | -1 | -1 | -2 | -2 | -1 | -1 | 1 | 0 |
| 5 | -2 | -1 | -1 | -1 | | 1 | 0 | -4 | -2 | -1 | -1 | -2 | -2 | -2 | 0 | -4 | -2 | -1 | -1 | 1 |
| 6 | -1 | 0 | 2 | 1 | 1 | | -1 | -3 | -1 | -2 | -2 | -1 | -1 | -2 | -1 | 1 | -3 | -3 | -1 | 0 |
| 7 | -1 | 0 | -2 | -1 | 0 | -1 | | -4 | 0 | -2 | 1 | -2 | -2 | -2 | -2 | 0 | 0 | -4 | -2 | -2 |
| 8 | 0 | -3 | -3 | -3 | -4 | -3 | -4 | | -1 | -2 | -2 | 3 | -3 | -2 | -1 | 0 | -1 | 0 | -3 | -2 |
| 9 | -2 | 0 | -1 | -1 | -2 | -1 | 0 | -1 | | -2 | -3 | -1 | -4 | -3 | 0 | -2 | -2 | -1 | 0 | -3 |
| 10 | 1 | 0 | 1 | -1 | -1 | -2 | -2 | -2 | -2 | | -3 | -1 | -3 | -1 | -3 | -2 | -1 | -1 | -1 | -1 |
| 11 | -3 | -2 | -2 | -3 | -1 | -2 | 1 | -2 | -3 | -3 | | -1 | -2 | -3 | -3 | -3 | -2 | -2 | -2 | -1 |
| 12 | 2 | -2 | -2 | -3 | -2 | -1 | -2 | 3 | -1 | -1 | -1 | | -3 | -3 | -2 | -3 | -4 | -3 | -1 | -2 |
| 13 | -3 | -4 | -1 | -1 | -2 | -1 | -2 | -3 | -4 | -3 | -2 | -3 | | -3 | -2 | -3 | -3 | -3 | -1 | -2 |
| 14 | -2 | 2 | -2 | -1 | -2 | -2 | -2 | -2 | -3 | -1 | -3 | -3 | -3 | | -3 | -3 | -3 | -3 | -3 | -1 |
| 15 | 0 | -1 | -1 | -2 | 0 | -1 | -2 | -1 | 0 | -3 | -3 | -2 | -2 | -3 | | -3 | -2 | -4 | -3 | -3 |
| 16 | 0 | 1 | -2 | -2 | -4 | 1 | 0 | 0 | -2 | -2 | -3 | -3 | -3 | -3 | -3 | | -3 | -3 | -2 | -3 |
| 17 | 0 | 2 | 0 | -1 | -2 | -3 | 0 | -1 | -2 | -1 | -2 | -4 | -3 | -3 | -2 | -3 | | -3 | -2 | -2 |
| 18 | 2 | 1 | 0 | -1 | -1 | -3 | -4 | 0 | -1 | -1 | -2 | -3 | -3 | -3 | -4 | -3 | -3 | | -3 | -3 |
| 19 | -1 | 1 | 0 | 1 | -1 | -1 | -2 | -3 | 0 | -1 | -2 | -1 | -1 | -3 | -3 | -2 | -2 | -2 | | -3 |
| 20 | 2 | -2 | 0 | 0 | 1 | 0 | -2 | -2 | -3 | -1 | -1 | -2 | -2 | -1 | -3 | -3 | -3 | -2 | -3 | |